\documentstyle[preprint,aps,epsfig]{revtex}
\preprint{CERN-TH/2000-026, \\ IMSc/2000/01/01, \\ hep-ph/0001310}
\def\be{\begin{equation}}
\def\ee{\end{equation}}

\def\ba{\begin{array}}
\def\ea{\end{array}}
\def\bea{\begin{eqnarray}}
\def\eea{\end{eqnarray}}

\def\l{\left}
\def\r{\right}

\def\lsim{\raise0.3ex\hbox{$\;<$\kern-0.75em\raise-1.1ex\hbox{$\sim\;$}}}
\def\gsim{\raise0.3ex\hbox{$\;>$\kern-0.75em\raise-1.1ex\hbox{$\sim\;$}}}

\begin{document}

\rightline{CERN-TH/2000-026}
\rightline{IMSc/2000/01/01}
\rightline{UMD-PP-00-053}
\rightline{hep-ph/0001310}

\begin{center}
{\large \bf Generation of large flavor mixing from
radiative corrections}

\medskip

{K. R. S. Balaji\footnote{balaji@imsc.ernet.in}}

{\it Institute of Mathematical Sciences, Chennai 600 113, India.}

\smallskip

{Amol S. Dighe\footnote{Amol.Dighe@cern.ch}}

{\it Div. TH, CERN, CH-1211 Geneva 23, Switzerland.}

\smallskip

{R. N. Mohapatra\footnote{rmohapat@physics.umd.edu}}

{ \it Department of Physics, University of Maryland, 
College park MD 20742, USA.} 

\smallskip

{M. K. Parida\footnote{mparida@vsnl.com}}

{\it Department of Physics, North Eastern Hill University, Shillong
793022, India.} 
\date{\today}

\end{center}

\begin{abstract}
We provide a model independent criterion
which would guarantee 
a large flavor mixing of two quasi-degenerate 
Majorana neutrinos 
at the low scale, irrespective of the mixing
at the high scale.
 We also show that such a situation is
realizable for a phenomenologically interesting range of parameters of the
weak scale theory. We further claim that for a similar condition to be 
implementable for the three generation case, 
the $CP$ parity of one of the neutrinos needs to
be opposite to that of the others.
\end{abstract}
\vspace{2cm}
\leftline{PACS numbers: 12.15.Lk, 14.60.Pq, 12.60.-i} 

\newpage

\section{INTRODUCTION}

The observed atmospheric and solar neutrino
 anomalies from the terrestrial experiments \cite{sk,solar} 
seem to provide a
strong evidence in favor  of neutrino flavor conversions,
implying that neutrinos are massive and they mix among
 themselves. While the details are fuzzy at this stage, 
it is clear that the atmospheric neutrino data require 
a large mixing of $\nu_\mu \leftrightarrow \nu_\tau$ or 
$\nu_\mu \leftrightarrow \nu_s$, although the latter possibility is
beginning to look less and less likely \cite{kajita}.
Regarding the solar neutrino anomaly, the mixing angle
(between $\nu_e$ and $\nu_{\mu,\tau}$ or $\nu_s$)
could be small, as in the small angle MSW scenario, 
or large, as in the large angle MSW or
vacuum oscillation scenarios \cite{solar}.

Since large mixing angles are involved in the possible solutions
of both the anomalies, a great deal of
theoretical work has gone in the understanding of the maximal mixing 
\cite{maximal}. There are two complementary 
approaches: (i) searching
for scenarios and symmetries beyond the standard model, and (ii) 
establishing general model independent criteria which guarantee the
stability at the weak scale of the masses and mixing pattern that
emerge at the high scale. The second approach has the advantage that
it may not only narrow down the search for new physics scenarios to a
manageable level, but it may also throw light on 
the parameters of the theory at the high scale, and on the
value of the high scale itself.
In this paper, we present some model independent criteria for such
theories, first focussing on the two flavor case and subsequently
on the three flavor models.

If neutrinos contribute
even a small fraction of the dark matter \cite{silk} of the universe, 
the oscillation observations imply a
situation where at least two neutrinos (and possibly even three) are
quasi-degenerate in mass. If the neutrino mass hierarchy is
{\it inverted} ($m_3 < m_1, m_2$ where $|\Delta m^2_{32}| \gg
|\Delta m^2_{21}|$), the neutrinos $\nu_1$ and $\nu_2$ are
necessarily quasi-degenerate. The study of scenarios where 
two or even three neutrinos are nearly degenerate is therefore
of crucial importance. 

The two flavor quasi-degenerate neutrino scenarios fall in two
classes: the neutrino flavors in the degenerate limit can be 
(i) in the same $CP$ eigenstate or (ii) in opposite $CP$ eigenstates.
It turns out that not only does one need to invoke different kinds of
symmetries to understand the two cases, but the radiative corrections
to the tree level degeneracy at the high scale can have very different
implications for the two cases. For instance, it has been noted in
\cite{haba,ema,ellis} that in the case (ii), the radiative
corrections (such as those through the RGE evolution from the
seesaw to the weak scale)
do not substantially affect the maximal mixing and 
quasi-degeneracy predicted by the theory at a high scale.

In this analysis, we point out that
in the case (i), an arbitrary mixing at the high scale
can get
``magnified'' to a large mixing, and even possibly
maximal mixing, at the low scale.
We find this interesting because 
(a) it enables a model builder to avoid any fine-tuning for the values of
mixing angles at the high scale, and hence
relaxes the constraints on the parameters of the high scale physics,
(b) it brings a certain unity in the understanding of the
quark and lepton mixings. This is arrived at by  
relating the radiative
corrections and the degree of mass degeneracy ($\frac{\delta m}{m}$),
regardless of the mixing pattern at
the high scale. 
In the context of specific models, this also leads to relationships
between the degree of degeneracy, the value of the high scale and 
the model parameters ({\it e.g.} $\tan \beta$ for the MSSM).
We further extend the results to the three generation scenario and
find that the constraint on $U_{e3}$ from the CHOOZ experiment 
\cite{chz1} indicates that the $CP$ parity of one of the neutrinos 
must be opposite to that of the others for our scheme to be implementable. 

Our paper is organized in the following form: in the next section, we
derive our main result for the two flavor mixing. In section III, we
present implications of the two flavor result for the case of the
standard model and the MSSM. In section IV, we
consider the extension to three generation case.

\section {BASIC FORMALISM FOR TWO FLAVORS}

Consider the mixing of two neutrinos.
The $2\times2$ Majorana matrix in the mass basis is of the form
\be
\cal{M_D}~=~\l(\ba{cc} m_{1} & 0 \\ 0 & m_{2} \ea \r).
\label{md}
\ee
The unitary matrix which takes $\cal{M_D}$ to the flavor basis
can be written as 
\be
U=\l(\ba{cc} C_\theta & S_\theta \\ -S_\theta & C_\theta \ea \r)
  \l(\ba{cc} 1 & 0 \\ 0 & e^{i\phi/2} \ea\r),  
\label{u-mat}
\ee
where $\theta$ is the mixing angle and $\phi$ is the $CP$
phase. All the quantities are defined at the high scale $\Lambda$.
The two neutrino flavors are related to
the mass eigenstates in the conventional form:
\be
\nu_f~=~ U_{fi}\nu_i, ~~~~~~~~~~~~~~~~~f~=~ {\alpha,\beta}~;
~i~=~{1,2}.
\label{u-def}
\ee  
We define the convention for the ``numbering'' of $\nu_1$ and
$\nu_2$ as follows. Let $\nu_\alpha$ and $\nu_\beta$
be the $SU(2)_L$ partners of the charged leptons
$\ell_\alpha$ and $\ell_\beta$ respectively, such that
$m_{\ell_\alpha} < m_{\ell_\beta}$. Then we define
 $\nu_1$ ($\nu_2$) as
the state with a larger component of the flavor $\alpha$ ($\beta$)
at the high scale. With this convention, $0 \le \theta \le \pi/4$.

The mass matrix in the flavor basis can be written as 
\bea
\cal{M_F} & = & U^* {\cal{M_D}} U^{\dag} \nonumber \\
  & = & \l(\ba{cc} C_\theta & S_\theta \\ -S_\theta & C_\theta \ea \r)
       \l(\ba{cc} m_1 & 0 \\ 0 & m_2 e^{-i\phi} \ea \r) 
        \l(\ba{cc} C_\theta & -S_\theta \\ S_\theta & C_\theta \ea \r).
\label{uudag}
\eea
 Let us examine the situation when $\phi=0$, which corresponds to the
case when the neutrinos $\nu_1$ and $\nu_2$ are in 
the same $CP$ eigenstate. 
Due to the presence of radiative corrections to $m_1$ and $m_2$,
the matrix $\cal{M_F}$ gets modified to
\be
\cal{M_F} \to \l(\ba{cc} 1+\delta_\alpha & 0 \\ 0 & 
1+\delta_\beta \ea \r) \cal{M_F} 
     \l(\ba{cc} 1+\delta_\alpha & 0 \\ 0 & 1+\delta_\beta \ea \r).
\label{mf-fin}
\ee
 In the above,
$\delta_\alpha$ and $\delta_\beta$ denote the corrections to the
masses in the flavor basis. The above general structure for $\cal M_F$
has been motivated by the RGE structure for radiative corrections \cite{babu}.
 We define
\be
\epsilon~\equiv~2(\delta_\beta - \delta_\alpha)~~,
\label{eps-def}
\ee
which is the net difference in the radiative
corrections for the masses of the two neutrino flavors.

 The mixing angle $\bar\theta$ that now diagonalizes the matrix
$\cal{M_F}$ at the low scale $\mu$ (after radiative corrections) 
can be related to the old mixing angle $\theta$
through the following expression:
\be
\tan 2\bar\theta~=~\tan 2\theta~(1~+~\delta_\alpha + \delta_\beta)
~\frac{1}{\lambda}~,
\label{t2tbar}
\ee
where
\be
\lambda \equiv
\frac{(m_2~-~m_1)C_{2\theta}~+~2\delta_\beta(m_1S_\theta^2~+~
m_2C_\theta^2)~-~2\delta_\alpha(m_1C_\theta^2~+~
m_2S_\theta^2)}{(m_2~-~m_1)C_{2\theta}}~.
\label{lam}
\ee
In the case of near degeneracy: $m_1\approx m_2 \approx m$, we have
\be
\lambda~=~\frac{m\epsilon }{(m_2~-~m_1)C_{2\theta}}~ + 1 ~,
\label{lam-1}
\ee
where $m$ is the common mass scale of the neutrinos. 

If 
\be
|m \epsilon| \gg |(m_2 - m_1) C_{2\theta}|~~,
\label{m-eps}
\ee
then 
$\lambda \rightarrow \infty$ and
we have $\tan2\bar\theta \to 0 $. 
Under this condition, any mixing angle tends to
zero after radiative corrections, i.e. a large mixing is unstable under
radiative corrections.
Note that this is true only for the two neutrinos with the same
$CP$ parity. If they had different $CP$ parities, {\it i.e.}
$\phi=\pi$, quasi-degeneracy would imply $|m_1| \approx |m_2| \approx m$,
however $|m_1 - m_2| \approx 2 m$. Then the radiative corrections
(which are small) cannot give the inequality (\ref{m-eps}).
In this case, $|m \epsilon| \ll |(m_2 - m_1) C_{2\theta}|$,
so that $\lambda \approx 1$ and the mixing angle does not change much. 
The mixing at the high scale then remains stable.
This reproduces the observations made in \cite{haba,ema} regarding
the stability  of the Maki-Nakagawa-Sakata (MNS) mixing matrix
\cite{mns}, when the  mixing angle is close to $\pi/4$. In addition, 
our analysis shows that the same conclusions remain valid for any arbitrary
nonzero $\theta$ of the MNS matrix.

If
\be
(m_1 - m_2)~ C_{2\theta}~=~2\delta_\beta(m_1S_\theta^2~+~
m_2C_\theta^2)~-~2\delta_\alpha(m_1C_\theta^2~+~
m_2S_\theta^2)~,
\label{cond}
\ee
then $\lambda = 0$ or equivalently
$\bar\theta = \pi/4$; {\it i.e.} maximal mixing. 
Given the mass heirarchy of the charged leptons: $m_{l_\alpha} \ll
m_{l_\beta}$, we expect $|\delta_\alpha| \ll |\delta_\beta|$,
which reduces (\ref{cond}) to a simpler form:
\be
\epsilon~=~\frac{\delta m C_{2\theta}}{(m_1S_\theta^2~+~
m_2C_\theta^2)}~,
\label{eps-simp}
\ee
where $\delta m \equiv m_1 - m_2$.
In the quasi-degenerate case,
\be
\epsilon ~\approx~ \frac{\delta m}{m} C_{2\theta}~.
\label{eps-dm}
\ee
The above expression can be translated in terms of the mass-squared
difference (which is the quantity measured in the oscillation
experiments) as 
\be
\epsilon~\approx~\frac{\Delta m^2(\Lambda)}{2m^2}C_{2\theta}~,
\label{eps-dmsq}
\ee
where $\Delta m^2(\Lambda)~=~ m_1^2(\Lambda)-m_2^2(\Lambda)$.
If the condition (\ref{eps-dmsq}) is satisfied,
the mixing at the scale $\mu$ tends to become maximal
regardless of the value of the mixing at the scale $\Lambda$.

Several points are worth emphasizing here. 
\\
1. The above relation between $\epsilon$ and the neutrino parameters
$\theta$, $m_1$ and $m_2$ is a model independent result and has
profound implications for model building. For instance, it will relax
the domain of parameters of the high scale theory compared to what was
believed earlier for $\phi =0$. 
\\
2. From (\ref{eps-dmsq}), 
the sign of $\epsilon$ must be the same as that of
$\Delta m^2(\Lambda)$ for getting maximal mixing at the low scale.
This preference is of a  phenomenological importance
since the sign of $\Delta m^2$ at low scales is measured by
experiments: if the solar neutrino solution is MSW, the 
identity of the heavier neutrino
is known, and the heavy / light nature of the third neutrino
may be determined through the long baseline experiments 
\cite{barger} or the observations of a galactic supernova \cite{amol}.
The model needs to be able to reproduce this sign from the
values of the masses at the high scale through the RGE. 
The results of this paper can thus be used to
discriminate between various models for a large flavor mixing. 
\\
3. The condition (\ref{eps-dmsq}) is not to be mistaken for 
a fine-tuning.
Though maximal mixing at the low scale requires an exact
equality (\ref{eps-dmsq}), the condition can be slackened if 
we only need a large mixing. Indeed, the SK data \cite{sk} 
indicate $|\tan 2\bar{\theta}| > 2$ at 90\% c.l..
In Fig. 1, we show the range of $\epsilon$ 
that allows a large mixing at the scale
$\mu$ as a function of the mixing angle at the scale $\Lambda$.
The region enclosed within the ``leaf'' gives the range of
$\epsilon$ which generates a large mixing 
($|\tan 2 \bar{\theta}| > 2$). 
The value of the degree of degeneracy $|\frac{\delta m}{m}|$ chosen
for the figure is 0.1. From (\ref{eps-dm}), changing this value 
would just change the scale of $\epsilon$ 
by a factor proportional to $|\frac{\delta m}{m}|$. 
The figure shows that a large mixing at
the scale $\mu$ is indeed 
possible for a large range of neutrino parameters.
The condition on the signs of $\epsilon$ and $\Delta m^2$ is 
also relaxed if the mixing at the high scale is 
already large.

\section {APPLICATIONS  TO THE STANDARD MODEL AND THE MSSM}

In this section, 
we analyze the implications of (\ref{eps-dmsq}) for the case
of the standard model (SM) and MSSM to see whether it is satisfied for
acceptable values of the model parameters.
In the case of the SM, the value of 
$\epsilon_{SM}$ from the RGE evolution is
\be
\epsilon_{SM} ~\approx~ \frac{h_\beta^2}{32 \pi ^2}ln(\frac{\Lambda}{M_Z}),
\label{eps-sm}
\ee
where $h_\beta$ corresponds to the Yukawa coupling of the 
heavier charged lepton.
Eq. (\ref{eps-dm}) and  the sign of $\epsilon_{SM}$ in (\ref{eps-sm})
imply that
for large flavor mixing to be generated in the SM through
radiative corrections, we require $m_1 >m_2$. 
In addition, from (\ref{eps-dmsq}) and (\ref{eps-sm}),
the strength of $h_\beta$ needs to be 
\be
h_\beta(SM) ~\approx~ \sqrt{\frac{16 \pi^2|\Delta m^2(\Lambda)|
C_{2\theta}}
{ln(\frac{\Lambda}{M_Z})m^2}}~.
\label{hbeta-sm}
\ee
This is a relation between $\Delta m^2(\Lambda)$ and 
the scale $\Lambda$ that needs to be obeyed. As an illustration, 
taking $\beta \equiv \mu$ (for $\nu_e \leftrightarrow \nu_\mu$, for
example), with $h_\mu \approx 6 \times 10^{-4}$
and the high scale as $\Lambda \sim 10^{12}$ GeV, for
degenerate neutrino mass of $m \sim 1$ eV, we get 
$|\Delta m^2(\Lambda)| \approx 10^{-7}$ eV$^2$.

In the case of MSSM, we have 
\be
\epsilon_{MSSM} ~\approx~ -\frac{h_\beta^2}{16 \pi ^2}ln(\frac{\Lambda}{\mu}).
\label{eps-mssm}
\ee
Eq. (\ref{eps-dm}) and the sign of $\epsilon_{MSSM}$ in 
(\ref{eps-mssm}) imply that we need $m_1 < m_2$ for large flavor
mixing to be generated through radiative corrections in the MSSM. 
In addition, from (\ref{eps-dmsq}) and (\ref{eps-mssm}), the
strength of $h_\beta$ has to be of
the order of 
\be
h_\beta (MSSM) ~\approx~ \sqrt{\frac{8 \pi^2|\Delta m^2(\Lambda)|
C_{2\theta}}
{ln(\frac{\Lambda}{\mu})m^2}}~.
\label{hbeta}
\ee
Taking $\beta \equiv \tau$ (for $\nu_\mu \leftrightarrow \nu_\tau$
mixing, for example), and using 
\be
h_\tau \approx \frac{m_\tau}{v \cos\beta}~,
\label{htau}
\ee
we get a relation between $\Lambda$, $\tan\beta$ and $m$, the
common mass scale of the neutrinos. 
In Fig. 2 we show the variation of $\epsilon$ with $\Lambda$
in MSSM for different values of  $h_\tau$. From 
(\ref{hbeta}) and (\ref{htau}), for 
given $m$ and $\Lambda$, 
we can infer the desirable value of $h_\tau$ and
hence of $\tan\beta$. For example, for $m \sim 1eV$ and
$\Lambda \sim 10^{12} GeV$, 
taking $|\Delta m^2 (\Lambda)| \approx 10^{-3} eV^2$,
we get $\tan\beta \approx 5$.

\section{Extension to three generations}

Let us now make a few comments on the possible extension 
to the case of three quasi-degenerate Majorana neutrinos.
If $m_{\alpha \beta}$ are the elements of the neutrino mass
matrix in the flavor basis, then in the approximation of
the decoupling of the third flavor, a large mixing
between flavors $\alpha$ and $\beta$ is guaranteed at the 
low scale if
\be
\epsilon_{\alpha \beta} \equiv 2(\delta_\beta - \delta_\alpha)
\approx 
\frac{2 (m_{\alpha \alpha} - m_{\beta \beta})
(1+\delta_\alpha + \delta_\beta)}{(m_{\alpha \alpha} + m_{\beta \beta})}~,
\label{eps-ab}
\ee
where no summation over repeated indices is implied.
Assuming that the $\Delta m^2$ hierarchy observed at the low scale is
true at the high scale also (small radiative corrections),
we have 
$|m_1 - m_2| \ll |m_2 - m_3|$.
The condition for $U_{\mu 3}$ to be maximal is then
\be
\epsilon_{\mu \tau} \approx \frac{(m_{3} - m_{2})(|U_{\tau 3}|^2 - 
|U_{\tau 2}|^2)}{m}.
\label{3-fl}
\ee
In all the models in which $h_\tau$ dominates over $h_e$ and $h_\mu$,
we have  $\epsilon_{e \tau} \approx \epsilon_{\mu \tau}$.
Then, the condition for the enhancement of $U_{\mu 3}$ 
(\ref{3-fl}) is similar to the condition for the enhancement of
$U_{e 3}$ (with the replacement $[\mu \to e, 2 \to 1]$) if all
the neutrinos have the same $CP$ parity,
assuming that both $U_{\tau 1},U_{\tau 2} \ll U_{\tau 3}$.
That would imply that when $U_{\mu 3}$ is magnified due to
radiative corrections, so is $U_{e 3}$.
Then one cannot naturally get a small
value of $U_{e3}$ at the low scale, as is 
suggested by the CHOOZ data \cite{chz1}.
Thus, in the three generation case with quasi-degenerate Majorana neutrinos,
we need the $CP$ phase of one neutrino opposite to that of the other
two in order for our condition to be implementable. It should be noted
that satisfying this condition still does not
guarantee the stability of small $U_{e3}$.
\\

In conclusion, we have derived a model independent condition that guarantees
a large mixing at the low scale irrespective of the 
mixing angle at the high scale, 
for two quasi-degenerate Majorana neutrinos with the same $CP$
parity. 
The condition relates the masses at the
high scale to the radiative corrections. 
In the case of SM and MSSM, this predicts 
the sign of the mass difference between the neutrinos and also
gives a range for its  magnitude. In
MSSM, it translates into a relation between the value of 
the high scale $\Lambda$, $\tan
\beta$, and the common mass of the neutrinos. 
Extending the argument to three quasi-degenerate Majorana 
neutrinos, we again show in a model independent way that 
the CP parity of one
of the neutrinos should be opposite to that of the others for our
conditions to be implementable at the phenomenological level.

\begin{center}
{\bf Acknowledgements}
\end{center}
We thank WHEPP-6, Chennai, India, where a part of the work was completed.
The work of RNM is supported by the NSF Grant no. PHY-9802551. The
work of MKP is supproted by 
the project No. 98/37/9/BRNS-cell/731 of the Govt. of India.

\begin{figure}
\vspace{-0.6in}
\epsfig{file=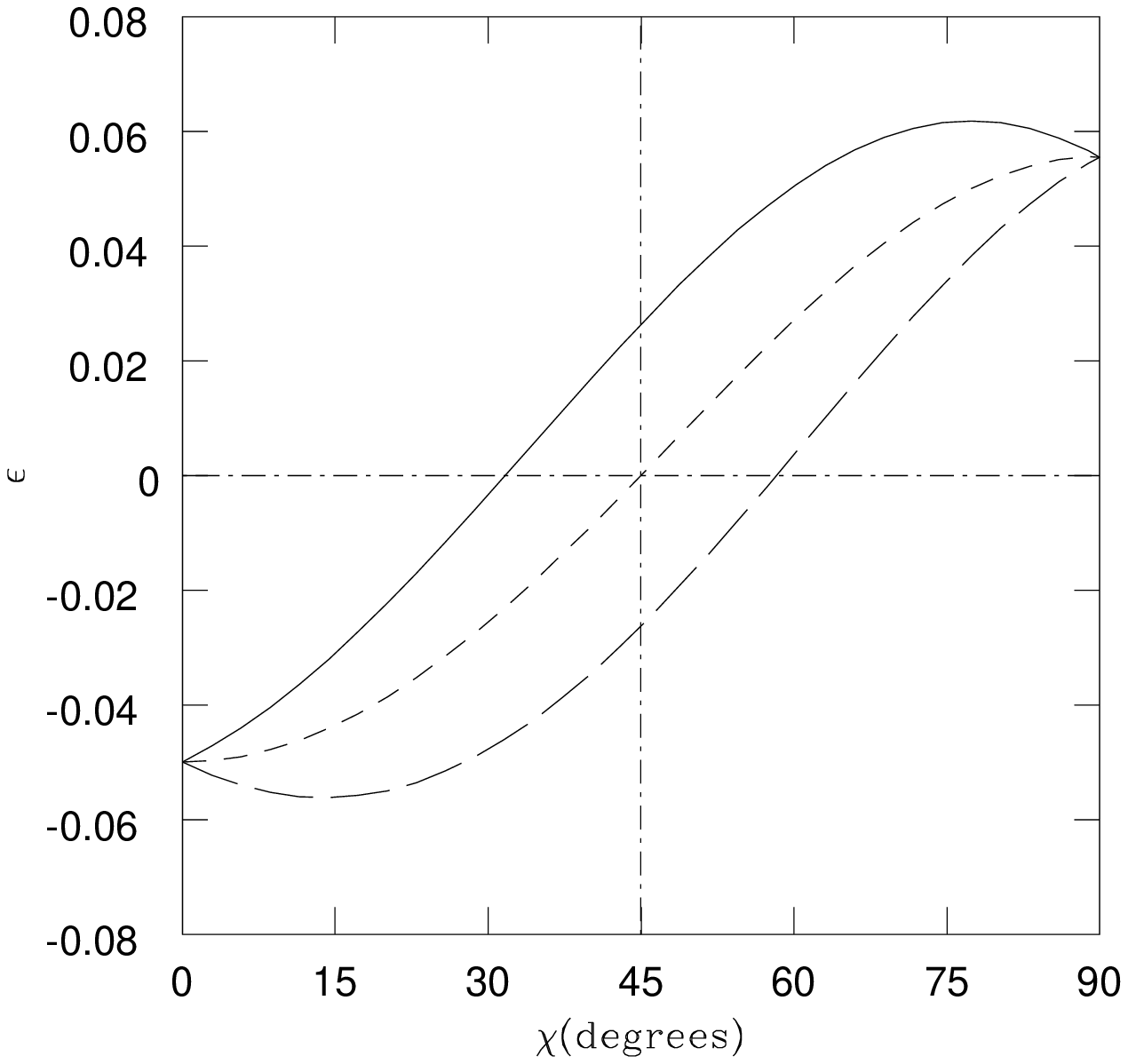,width=7in}
\caption{ The radiative correction $\epsilon$ that can generate a large
mixing at the scale $\mu$ for $|\frac{\delta m}{ m}|=0.1$. 
Here $\chi \equiv \theta$ for $m_1 < m_2$ and 
$\chi \equiv \frac{\pi}{2} - \theta$ for $m_1 > m_2$.
The central short-dashed
line corresponds to maximal mixing ($\bar{\theta} = \frac{\pi}{4}$).
The solid (long-dashed) line denotes $\tan 2\bar{\theta} = 2 (-2)$.
}
\label{leaf}
\end{figure}

\begin{figure}
\vspace{-0.6in}
\epsfig{file=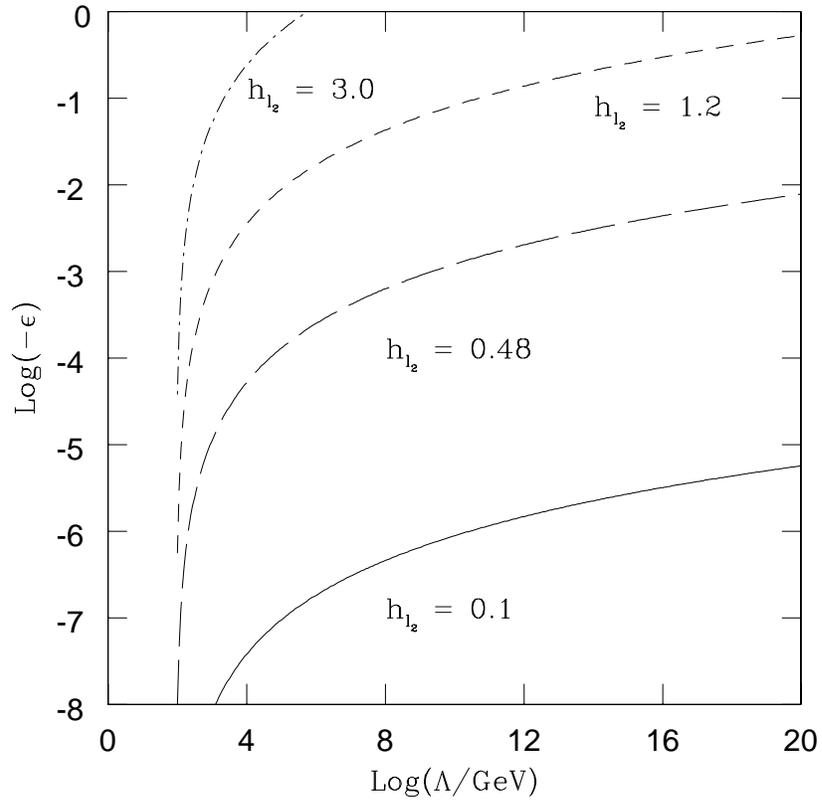,width=7in}
\vspace{-0.6in}
\caption{Log-Log plot of $\epsilon$ vs $\Lambda$ for a few values of
$h_{l_2} \equiv h_\beta$ in the MSSM.}
\label{plot}
\end{figure}

\end{document}